\documentclass[12pt]{article}

\usepackage[dvips]{graphics}

\def\beq   {\begin{equation}}
\def\eeq   {\end{equation}}
\def\beqd  {\begin{displaymath}}
\def\eeqd  {\end{displaymath}}
\def\beqaa {\begin{eqnarray}}
\def\eeqaa {\end{eqnarray}}

\def\noi {\noindent}

\def\ti  {\tilde}

\def\st  {\ti t}
\def\sb  {\ti b}
\def\sg  {\ti g}
\def\sa  {\ti \tau}
\def\sn  {\ti \nu}
\def\sl  {\ti \ell}

\def\nt  {\tilde\chi^0}
\def\ch  {\tilde\chi^\pm}
\def\chp {\tilde\chi^+}
\def\chm {\tilde\chi^-}

\def\a   {\alpha}
\def\b   {\beta}
\def\t   {\theta}

\def\tsa {\theta_{\sa}}

\def\sz{\ifmmode{\tilde{\chi}^0} \else{$\tilde{\chi}^0$} \fi}
\def\sw{\ifmmode{\tilde{\chi}} \else{$\tilde{\chi}$} \fi}

\newcommand{\gsim}{\;\raisebox{-0.9ex}
           {$\textstyle\stackrel{\textstyle >}{\sim}$}\;}
\newcommand{\lsim}{\;\raisebox{-0.9ex}{$\textstyle\stackrel{\textstyle<}
           {\sim}$}\;}

\begin{document}
\pagestyle{empty}

\vspace*{-1cm} 
\begin{flushright}
  UWThPh-1999-21 \\
  HEPHY-PUB 713/99 \\
  TGU-24 \\
  AP-SU-99/01 \\
  TU-561 \\
  hep-ph/9904417
\end{flushright}

\vspace*{1.4cm}

\begin{center}

{\Large {\bf
Impact of bosonic decays on the search for 
\boldmath{$\tilde \tau_2$} and \boldmath{$\tilde \nu_\tau$}
}}\\

\vspace{10mm}

{\large 
A.~Bartl$^a$, H.~Eberl$^b$, K.~Hidaka$^c$, S.~Kraml$^b$, 
T.~Kon$^d$,\\[2mm]
W.~Majerotto$^b$, W.~Porod$^a$, and Y.~Yamada$^e$}

\vspace{6mm}

\begin{tabular}{l}
$^a${\it Institut f\"ur Theoretische Physik, Universit\"at Wien, A-1090
Vienna, Austria}\\
$^b${\it Institut f\"ur Hochenergiephysik der \"Osterreichischen Akademie
der Wissenschaften,}\\
\hphantom{$^b$}{\it A-1050 Vienna, Austria}\\
$^c${\it Department of Physics, Tokyo Gakugei University, Koganei,
Tokyo 184--8501, Japan}\\
$^d${\it Faculty of Engineering, Seikei University, Musashino, Tokyo 180--8633,
Japan}\\
$^e${\it Department of Physics, Tohoku University, Sendai 980--8578, Japan}
\end{tabular}

\end{center}

\vfill

\begin{abstract} 
We perform a detailed study of the decays of the heavier $\tau$ slepton 
($\sa_2$) and $\tau$-sneutrino ($\sn_\tau$) 
in the Minimal Supersymmetric Standard Model (MSSM). 
We show that the decays into Higgs or gauge bosons, i.e. 
$\sa^-_2 \to \sa^-_1 + (h^0, H^0, A^0 \ \mbox{or} \ Z^0)$,  
$\sa^-_2 \to \sn_{\tau} + (H^- \ \mbox{or} \ W^-)$, and 
$\sn_\tau \to \sa^-_1 + (H^+ \ \mbox{or} \ W^+)$, 
can be very important 
due to the sizable $\tau$ Yukawa coupling and large mixing parameters 
of $\sa$. 
Compared to the decays into fermions, such as 
$\sa^-_2 \to \tau^- + \nt_i$ and $\sa^-_2 \to \nu_\tau + \chm_j$, 
these bosonic decay modes can have significantly different decay 
distributions. 
This could have an important influence on the search for $\sa_2$ 
and $\sn_\tau$ and the determination of the MSSM parameters 
at future colliders.
\end{abstract}

\newpage
\pagestyle{plain}
\setcounter{page}{2}


In the Minimal Supersymmetric Standard Model (MSSM) \cite{ref1} 
supersymmetric (SUSY) partners of all Standard Model 
(SM) particles with masses less than O(1 TeV) are introduced. 
This solves the problems of hierarchy, fine-tuning and 
naturalness of the SM. 
Hence discovery of all SUSY partners and 
study of their properties are essential for testing the MSSM.  
Future colliders, such as the Large Hadron Collider (LHC), the upgraded 
Tevatron, $e^+e^-$ linear colliders, and $\mu^+\mu^-$ colliders will 
extend the discovery potential for SUSY particles to the TeV mass range 
and allow for a precise determination of the SUSY parameters.

\noi
In this article we focus on the sleptons of the third generation, i.~e. 
staus ($\sa_{1,2}$; $m_{\sa_1}<m_{\sa_2}$) and 
tau-sneutrino ($\sn_\tau$). These particles may have properties 
different from the sleptons of the other two generations 
due to the sizable $\tau$ Yukawa coupling. 
Production and decays of $\sa_i$ and $\sn_\tau$ were studied 
in \cite{ref3a,ref3,ref10}. 
Like other sleptons, they can decay into fermions, i.~e. 
a lepton plus a neutralino ($\nt_k$) or chargino ($\ch_j$): 
\beq
  \begin{array}{lcl}
  \sa^-_i \to \tau^-\, \nt_k\,, &\hspace{3mm}& 
  \sn_\tau \to \nu_\tau\,  \nt_k\,, \\
  \sa^-_i \to \nu_\tau\,  \chm_j\,, &\hspace{3mm}& 
  \sn_\tau \to \tau^-\,  \chp_j\,,
  \end{array}
  \label{eq:fmodes}
\eeq
with $i,j=1,2$ and $k=1,...,4$. In addition, the heavier stau $\sa_2$ 
and the tau-sneutrino can also decay into
bosons \cite{ref3,ref10}, i.~e. a lighter slepton plus a gauge boson
\beq
  \begin{array}{lcl}
    \sa^-_2 \to \sa^-_1\,Z^0\,, &  \\
    \sa^-_2 \to \sn_\tau\:W^-\,, &\hspace{3mm}& \sn_\tau \to \sa^-_1\:W^+\,, 
  \end{array}
  \label{eq:wzmodes}
\eeq
or a Higgs boson 
\beq
  \begin{array}{lcl}
    \sa^-_2 \to \sa^-_1\:(h^0,\,H^0,\,A^0)\,, & & 
      \\
    \sa^-_2 \to \sn_\tau\:H^-\,, &\hspace{3mm}& \sn_\tau \to \sa^-_1\:H^+\,.
  \end{array}
  \label{eq:hxmodes}
\eeq
The decays in Eqs.~(\ref{eq:wzmodes}) and (\ref{eq:hxmodes}) are 
possible in case the mass splitting between the sleptons is 
sufficiently large. 

\noi
In the present article we perform a more general analysis 
than \cite{ref3,ref10}. 
We point out that the $\sa^-_2$ and $\sn_\tau$ decays into gauge or Higgs 
bosons of Eqs.~(\ref{eq:wzmodes}) and (\ref{eq:hxmodes})
can be very important in a large region of the MSSM parameter space due to 
the sizable $\tau$ Yukawa coupling and 
large $\tilde{\tau}$ mixing parameters.  
This importance of the bosonic modes relative to the 
fermionic modes of Eq.~(\ref{eq:fmodes})
could have a significant influence on searches for 
$\sa^-_2$ and $\sn_\tau$ at future colliders.
An analogous study for the heavier stop ($\st_2$) and sbottom ($\sb_2$) 
was performed in \cite{ref4,ref5}.


First we summarize the MSSM parameters in our analysis. 
In the MSSM the stau sector is specified by the mass matrix in 
the basis $(\sa_L^{},\sa_R^{})$ \cite{ref6,ref7}
\begin{equation}
  {\cal M}^2_{\sa}= 
     \left( \begin{array}{cc} 
                m_{\sa_L}^2 & a_\tau m_\tau \\
                a_\tau m_\tau     & m_{\sa_R}^2
      \end{array} \right)       
  \label{eq:f}
\end{equation}
with
\begin{eqnarray}
  m_{\sa_L}^2 &=& M_{\ti L}^2 
                  + m_Z^2\cos 2\beta\,(\sin^2\t_W - \frac{1}{2}) 
                  + m_\tau^2, \label{eq:g} \\
  m_{\sa_R}^2 &=& M_{\ti E}^2  
                  - m_Z^2 \cos 2\b\, \sin^2\t_W + m_\tau^2, 
                                              \label{eq:h}\\[2mm]
  a_\tau m_\tau     &=& 
                     (A_\tau - \mu\tan\beta)\,m_\tau \, .
                   \label{eq:i}
\end{eqnarray}
$M_{\ti L,\ti E}$ and $A_\tau$ are soft SUSY--breaking 
parameters, $\mu$ is the higgsino mass parameter, 
and $\tan\b = v_2/v_1$ with $v_1$ $(v_2)$ being the vacuum 
expectation value of the Higgs field $H_1^0$ $(H_2^0)$. 
Diagonalizing the matrix (\ref{eq:f}) one gets the mass eigenstates 
$\sa_1=\sa_L\cos\t_{\sa}+\sa_R\sin\t_{\sa}$, and 
$\sa_2=-\sa_L\sin\t_{\sa}+\sa_R\cos\t_{\sa}$ 
with the masses $m_{\sa_1}$, $m_{\sa_2}$ ($m_{\sa_1}<m_{\sa_2}$) 
and the mixing angle $\t_{\sa}$. 
The stau mixing is large if 
$|m_{\sa_L}^2-m_{\sa_R}^2| \lsim |a_\tau m_\tau|$, which may be the case 
for large $\tan\beta$ and $\mu$. 
The mass of $\sn_\tau$ is given by 
\beq
m^2_{\sn_\tau} = M^2_{\ti L} + {\frac{1}{2}}m^2_Z\cos{2\beta}. 
\eeq
\\
The properties of the charginos $\ch_i$ ($i=1,2$; $m_{\ch_1}<m_{\ch_2}$) 
and neutralinos $\nt_k$ ($k=1,...,4$; $m_{\nt_1}< ...< m_{\nt_4}$)  
are determined by the parameters $M$, $M'$, $\mu$ and $\tan\b$, 
where $M$ and $M'$ are the SU(2) and U(1) gaugino masses, respectively. 
Assuming gaugino mass unification we take $M'=(5/3)\tan^2\t_W M$. 
The masses and couplings of the Higgs bosons $h^0, ~H^0, ~A^0$, 
and $H^{\pm}$, 
including leading radiative corrections, are fixed by 
$m_A,~\tan\beta,~\mu,~m_t,~m_b,~M_{\ti Q},~M_{\ti U},~M_{\ti D}, 
~A_t,$ and $A_b$, 
where $M_{\ti Q,\ti U,\ti D}$ and $A_{t, b}$ are soft SUSY-breaking 
parameters in the ($\st$, $\sb$) sector. 
$H^0$ ($h^0$) and $A^0$ are the heavier (lighter) CP--even and CP--odd 
neutral Higgs bosons, respectively. 
For the radiative corrections to the $h^0$ and $H^0$ masses 
and their mixing angle $\alpha$ we use the formulae of 
Ref.~\cite{ref8}; 
for those to $m_{H^+}$ we follow Ref.~\cite{ref9}.\footnote{
Notice that \cite{ref8,ref9} have a sign convention 
for the parameter $\mu$ opposite to the one used here. 
} 
We treat $M_{\{{\ti L},{\ti E},{\ti Q},{\ti U},{\ti D}\}}$ and 
$A_{\{{\tau},{t},{b}\}}$ as free parameters. 

\noi
The widths of the $\sl_i=\sa^-_2$ or $\sn_\tau$ decays into 
Higgs and gauge bosons are given by ($k=1,...,4$) \cite{ref10}:
\beq
  \Gamma (\sl_i \to \sl_j^{(\prime)}\,H_{\!k}^{}) = 
    \frac{\kappa_{ijk}^{}}{16\pi\,m_{\sl_i}^3}\;(G_{ijk})^2 ,
  \hspace{4mm}
  \Gamma (\sl_i \to \sl_j^{(\prime)}\,V) = 
    \frac{\kappa_{ijV}^3}{16\pi\,m_V^2\,m_{\sl_i}^3}\;(c_{ijV})^2.
\label{eq:widths}
\eeq
Here $\sl_j^{(\prime)} = \sa^-_1$ or $\sn_\tau$ 
(with the indices $i$ and $j$ to be omitted for $\sn_\tau$), 
$H_{\!k}^{}=\{h^0\!,\,H^0\!,\,A^0\!,\,H^\pm\}$ and $V=\{Z^0,W^\pm\}$. 
$\kappa_{ijX}^{}\equiv\kappa(m_{\sl_i}^2,m_{\sl_j^{(\prime)}}^2,m_X^2)$
is the usual kinematic factor, 
$\kappa(x,y,z)=(x^2+y^2+z^2-2xy-2xz-2yz)^{1/2}$. 
Notice an extra factor $\kappa^2/m^2_V$ for the gauge boson modes. 
$G_{ijk}$ denote the slepton couplings to Higgs bosons 
and $c_{ijV}$ those to gauge bosons.
Their complete expressions, as well as the widths 
of the fermionic decays, are given in \cite{ref10}. 

\noi
Since $m_\tau$ is rather small, we need a large difference between 
$m_{\sa_L}\sim m_{\sn_\tau}$ and $m_{\sa_R}$ in order to realize 
a mass splitting 
between ($\sa_1$, $\sa_2$, $\sn_\tau$) large enough to allow 
the bosonic decays in Eqs.~(2, 3). In this case the $\sa_L-\sa_R$ mixing 
is rather small. Thus, in this article we consider two patterns of 
the mass spectrum of the sleptons: 
$m_{\sa_1} < m_{\sa_2}\sim m_{\sn_\tau}$ with 
$(\sa_1,\sa_2)\sim(\sa_R,\sa_L)$ for $m_{\sa_L} > m_{\sa_R}$, 
and 
$m_{\sa_2} > m_{\sa_1}\sim m_{\sn_\tau}$ with 
$(\sa_1,\sa_2)\sim(\sa_L,\sa_R)$ for $m_{\sa_L} < m_{\sa_R}$.
The bosonic decays are therefore basically the decays of 
$(\sa_L , \sn_\tau)$ into $\sa_R$ or vice versa. 
This is in strong contrast to the case of the $\st_2$ and 
$\sb_2$ decays \cite{ref5}. In the latter case, large mass splittings can be 
obtained even for $M_{\ti Q}\sim M_{\ti U}\sim M_{\ti D}$ 
due to the large top and/or bottom Yukawa couplings, 
which can also cause large left-right mixings and a complex decay spectrum. 

\noi
The leading terms of $G_{ijk}$ and $c_{ijV}$ which are relevant for 
the bosonic decays are given in Table~1. Here the Yukawa 
coupling $h_{\tau}$ is given by 
\beq
  h_\tau = g\,m_\tau/(\sqrt{2}\,m_W\cos\b) \, .
\eeq
%
\begin{table}[ht] \hrule \vspace{1mm}
\beqd
  \begin{array}{lll}
    \sa_1\sa_2h^0 & \!\!\sim\!\!
                  & h_\tau\,(\mu\cos\a + A_\tau\sin\a) \cos 2\tsa 
    \\
    \sa_1\sa_2H^0 & \!\!\sim\!\!
                  & h_\tau\,(\mu\sin\a - A_\tau\cos\a) \cos 2\tsa 
    \\
    \sa_1\sa_2A^0 & \!\!\sim\!\!
                  & h_\tau\,(\mu\cos\b + A_\tau\sin\b)  
    \\
    \sa_1\sa_2Z^0 & \!\!\sim\!\! & g\,\sin 2\tsa 
  \end{array}
\eeqd
\beqd
  \begin{array}{lll}
     \sn_\tau\sa_1H^\pm &\!\!\sim\!\!
             & h_\tau\,(\mu\cos\b + A_\tau\sin\b)\sin\tsa \\
     \sn_\tau\sa_2H^\pm &\!\!\sim\!\!
             & h_\tau\,(\mu\cos\b + A_\tau\sin\b)\cos\tsa 
\\
     \sn_\tau\sa_1W^\pm &\!\!\sim\!\!
             & g \,\cos\tsa \\
     \sn_\tau\sa_2W^\pm &\!\!\sim\!\!
             & - g \,\sin\tsa 
  \end{array}   
\eeqd
\hrule
\caption{Leading terms of the slepton couplings to Higgs and gauge bosons.}
\label{table1}
\end{table} 
%
%
Higgs bosons couple mainly to 
$(\sa_L,\sn_\tau)-\sa_R$ combinations. 
These couplings are proportional to the Yukawa coupling 
$h_\tau$ and the parameters $A_\tau$ and $\mu$, 
as can be seen in Table~1. 
Notice the factor $\cos 2\t_{\sa}$ in the $\sa_1^{}\sa_2^{}h^0$ 
and $\sa_1^{}\sa_2^{}H^0$ couplings, but not in the 
$\sa_1^{}\sa_2^{}A^0$ coupling. 
In our case $|\cos 2\t_{\sa}| \sim 1$ due to the small $\sa_L-\sa_R$ mixing, 
unlike the case of $\st_2$ and $\sb_2$ decays \cite{ref5}. 
Similarly, the $\sn_\tau \sa_i H^\pm$ coupling is not suppressed 
by $\sa$ mixing for $\sa_i\sim\sa_R$. 
Hence the widths of the decays into Higgs bosons 
can be large for large $\tan\beta$, $A_\tau$, and $\mu$. 
Notice that the $\sa_L-\sa_R$ mixing enhances the 
mass splitting between $\sa_1$ and $\sa_2$, which results in 
a larger phase space for the bosonic decays of $\sa_2$. 
In contrast, the gauge couplings which are relevant for the bosonic decays 
are suppressed by the small mixing of sleptons, since the 
gauge interactions preserve the chirality of sleptons. 
However, this suppression is largely compensated by the extra factor 
$\kappa^2/m_V^2$ in Eq.~(\ref{eq:widths}). In fact, 
since the gauge bosons in the decays (\ref{eq:wzmodes}) 
are longitudinally polarized, the widths of the decays into gauge 
bosons for $m_{\sl_i} - m_{\sl_j^{(\prime)}} \gg m_V$ 
are approximated by those into the 
corresponding Nambu-Goldstone (NG) bosons. 
As a result, the decays into gauge bosons are enhanced 
when the couplings of the sleptons to the NG bosons 
$(\propto a_\tau m_\tau / m_V)$ are large.
On the other hand, the fermionic decays are not enhanced for large 
($A_\tau$, $\mu$, $\tan\beta$). Although the slepton-Higgsino couplings 
increase with $\tan\beta$, their contribution to the kinematically 
accessible decays (\ref{eq:fmodes}) is suppressed for $|\mu|>m_{\sa_2}$. 
Therefore the branching ratios of the bosonic decays 
(\ref{eq:wzmodes}, \ref{eq:hxmodes}) are expected to be large for 
large $A_\tau$, $\mu$, and $\tan\beta$.


We now turn to the numerical analysis of the $\sa_2$ and $\sn_\tau$ decay 
branching ratios. 
We calculate the widths of all possibly important 
two-body decay modes of Eqs. (\ref{eq:fmodes}, \ref{eq:wzmodes}, 
\ref{eq:hxmodes}). 
Three-body decays are negligible in this study.
We take  $m_\tau = 1.78$ GeV, 
$m_t=175$ GeV, $m_b=5$ GeV, $m_Z^{}=91.2$ GeV, $\sin^2\t_W =0.23$, 
$m_W^{} = m_Z^{}\cos\t_W$ and $\alpha(m_Z^{})=1/129$.
In order not to vary too many parameters we fix 
$M=300$ GeV, $m_A=150$ GeV, and 
$M_{\tilde{Q}}$ $=$ $M_{\tilde{U}}$ $=$ $M_{\tilde{D}}$ 
$=$ $A_t$ $=$ $A_b$ $=$ 500 GeV for simplicity. 
In our numerical study we take 
$m_{\sa_1}$, $m_{\sa_2}$, $A_\tau$, $\mu$, and $\tan\beta$ 
as input parameters. 
Note that 
for a given set of the input parameters we have two 
solutions for ($M_{\ti L}$, $M_{\ti E}$) corresponding to the 
two cases 
$m_{\sa_L} \geq m_{\sa_R}$ and $m_{\sa_L} < m_{\sa_R}$.
In the plots we impose the following conditions 
in order to respect experimental and theoretical constraints:
\renewcommand{\labelenumi}{(\roman{enumi})} 
\begin{enumerate}
  \item $m_{\ch_1} > 100$ GeV, $m_{\sn_\tau} > $ 45 GeV, 
        $m_{h^0} > $ 90 GeV, $m_{\sa_1,\st_1,\sb_1} > m_{\nt_1}$, 
        $m_{\nt_1} > 80$ GeV, 
  \item $A_\tau^2 < 3\,(M_{\ti L}^2 + M_{\ti E}^2 + m_{H_1}^2)$, 
        $A_t^2 < 3\,(M_{\ti Q}^2 + M_{\ti U}^2 + m_{H_2}^2)$, and 
        $A_b^2 < 3\,(M_{\ti Q}^2 + M_{\ti D}^2 + m_{H_1}^2)$, where 
        $m_{H_1}^2=(m_A^2+m_Z^2)\sin^2\b-\frac{1}{2}\,m_Z^2$ and 
        $m_{H_2}^2=(m_A^2+m_Z^2)\cos^2\b-\frac{1}{2}\,m_Z^2$, 
  \item $\Delta\rho\,(\st\!-\!\sb) < 0.0012$ 
        \cite{ref11} using the formula of \cite{ref12}.
\end{enumerate}
Condition (i) is imposed to satisfy the 
experimental bounds on $\tilde\chi_1^\pm$, $\tilde\chi_1^0$, $\sa$, $\sn$, 
$\st$, $\sb$, $\sg$, and $h^0$ from LEP \cite{ref14} and Tevatron 
\cite{ref15}. Note that $m_{\nt_1} > 80$ GeV is imposed in order to 
evade the experimental bounds on $m_{\st_1}$ and $m_{\sb_1}$.
Condition (ii) is approximately necessary to avoid 
color and charge breaking global minima \cite{ref13} and 
to exclude unrealistically large $A_{\tau}$. 
Condition (iii) constrains $\mu$ and $\tan\beta$ in the squark sector. 
We note that the experimental data for the $b \to s\gamma$ decay give 
rather strong constraints \cite{ref16} on the SUSY and Higgs parameters 
within the minimal supergravity model, especially for large $\tan\beta$. 
However, we do not impose  this constraint since it strongly depends on 
the detailed properties of the squarks, including the generation-mixing.

\noi
In Fig.~1 we plot in the $A_\tau$--$\mu$ plane the contours of 
the branching ratios of the Higgs boson modes 
${\rm BR}(\sl^{}\to\sl^{\prime}H)$ $\equiv$ 
$\sum {\rm BR}\big[\:\sl^{}\to \sl' + (h^0,H^0,A^0,H^\pm) \,\big]$, 
the gauge boson modes 
${\rm BR}(\sl^{}\to\sl^{\prime}V)$ $\equiv$ 
$\sum {\rm BR}\big[\:\sl^{}\to \sl' + (Z^0,W^\pm) \,\big]$, 
and the total bosonic modes 
${\rm BR}(\sl^{}\to\sl^{\prime}B)$ $\equiv$ 
${\rm BR}(\sl^{}\to\sl^{\prime}H)$ $+$ 
${\rm BR}(\sl^{}\to\sl^{\prime}V)$, 
with $\sl=(\sa_2^-$, $\sn_\tau)$ and 
$\sl'=(\sa_1^-$, $\sn_\tau)$. 
We take $m_{\sa_1} = 250$ GeV, $m_{\sa_2} = 500$ GeV, $\tan\beta = 30$ 
and show two cases $m_{\sa_L} < m_{\sa_R}$ and 
$m_{\sa_L} \geq m_{\sa_R}$. 
Note that $\sn_\tau$ decays into bosons and $\sa_2^-$ decays into 
$\sn_\tau$ are kinematically forbidden if 
$m_{\sa_L} < m_{\sa_R}$ and $m_{\sa_L} \geq m_{\sa_R}$, respectively. 

\noi
We observe that ${\rm BR}(\sl^{}\to\sl^{\prime}H)$ increases 
with $|A_\tau|$ while ${\rm BR}(\sl^{}\to\sl^{\prime}V)$ 
increases with $|\mu|$. This dependence on ($A_\tau$, $\mu$) 
is explained as follows. 
Note first that for large $\tan\beta$ the mixing between $H_1$ and 
$H_2$ is rather small. Hence, for $m_A > m_Z$, ($H^0$, $A^0$, $H^\pm$) 
are mainly $H_1$ while $h^0$ and the NG bosons are mainly $H_2$. 
The couplings of $(\sa,\sn_\tau)$ to $H_1$ and $H_2$ are $\sim$ 
$h_\tau A_\tau$ and $h_\tau\mu$, respectively. Therefore the decays 
to ($H^0$, $A^0$, $H^\pm$) are enhanced for large $|A_\tau|$, whereas 
those to $h^0$ and to gauge bosons are enhanced for large $|\mu|$. 
This property can also be derived directly from Table 1 by noting that
$|\sin\alpha| \ll 1$ and $\sin\beta \sim 1$. 
As a result, the total bosonic branching ratio 
${\rm BR}(\sl^{}\to\sl^{\prime}B)$ becomes large, and 
even dominant for $m_{\sa_L}<m_{\sa_R}$, in a 
wide region of the $A_\tau$--$\mu$ plane, especially 
for large $|A_\tau|$ and/or $|\mu|$, as seen in Fig.~1. 
We also see that the branching ratios of the bosonic decays are almost 
unchanged by $A_\tau$ $\to$ $-A_\tau$ and/or $\mu$ $\to$ $-\mu$. 

\noi
In Fig.~2 we show the individual branching ratios of 
the $\sa_2$ and $\sn_\tau$ decays 
as a function of $\tan\beta$ for 
$m_{\sa_1} = 250$ GeV, $m_{\sa_2} = 500$ GeV, 
$A_\tau = 800$ GeV 
and $\mu = 1000$ GeV, for the two cases 
$m_{\sa_L} \geq m_{\sa_R}$ and $m_{\sa_L} < m_{\sa_R}$. 
We see that the branching ratios of the boson modes 
increase with $\tan\beta$ and become important, and even dominant if 
$m_{\sa_L}<m_{\sa_R}$, for large 
$\tan\beta\, (\hspace{-1.4mm}\gsim \hspace{-1mm}15)$. 
As already explained, this comes from the increase of $h_\tau$ 
and $a_\tau m_\tau$ with $\tan\beta$. We have checked that 
$\Gamma (\sa_2 \to \sa_1 H^0)$ itself increases with $\tan\beta$.

\noi 
In Fig.~3 we show the $m_{\sa_2}$ dependence 
of the $\sa_2$ and $\sn_\tau$ decay branching ratios for 
$m_{\sa_1} = 250$ GeV, $A_\tau = 800$ GeV, $\mu = 1000$ GeV and 
$\tan\beta = 30$. 
We see that the branching ratios of the bosonic decays decrease 
with increasing $m_{\sa_2}$. This behavior comes from the fact that 
in the large $m_{\sa_2}$ limit the decay widths of the bosonic and 
fermionic modes are proportional to $m_{\sa_2}^{-1}$ and $m_{\sa_2}$, 
respectively. 

\noi
In all figures one can see that the total branching ratio of the 
bosonic decays of $\sa_2$ 
is substantially larger for $m_{\sa_L} < m_{\sa_R}$ than for 
$m_{\sa_L} \geq m_{\sa_R}$. 
The reason is the following: 
First, as already explained, the decays of $\sa_2^-$ into 
$\sn_\tau$ are kinematically allowed only for $m_{\sa_L} < m_{\sa_R}$, 
as seen in Figs.~2 and 3. 
Second, for the parameter set used in Figs.~2 and 3, 
$\chm_1 \sim {\ti W}^-$. 
Hence the decay $\sa_2^- \to \nu_\tau \chm_1$ is 
strongly suppressed for $m_{\sa_L} < m_{\sa_R}$ 
(i.e. $\sa_2 \sim \sa_R$), but not for $m_{\sa_L} \geq m_{\sa_R}$, 
as seen in Figs.~2 and 3; the $\sa_2$ decay into 
$\chm_2 \sim {\ti H}^-$ is kinematically forbidden.
This results in a rather strong enhancement of the bosonic modes 
for the case $m_{\sa_L} < m_{\sa_R}$. 

\noi
We find that the importance of the bosonic 
modes is fairly insensitive to the choice of the values of 
$m_A$, $M$, $M_{{\ti Q},{\ti U},{\ti D}}$ and $A_{t,b}$. 
The decays to $H^0$, $A^0$ and $H^\pm$ are kinematically 
suppressed for large $m_A$. 
However, the remaining $h^0$ and gauge boson modes can still be 
important and even dominant. 
For example, for $m_{\sa_1}=250$ GeV, $m_{\sa_2}=500$ GeV, 
$A_{\tau}=800$ GeV, $\mu=1000$ GeV, $\tan\beta = 30$ and 
$m_{\sa_L} < m_{\sa_R}$, 
the $\sa_2$ decays into ($H^0$, $A^0$, $H^\pm$) are forbidden for 
$m_A>260$ GeV. Nevertheless, in this case 
we have BR($\sa_2 \to \sa_1 h^0$) $=$ 13\% and 
BR($\sa_2 \to \sl V$) $=$ 52\%. 
We have also checked that our results do not 
change significantly for smaller values of $M$ in the range 
where the condition (i) is satisfied. 

\noi
Now we discuss the signatures of the $\sa_2$ and $\sn_\tau$ decays. 
We compare the signals of the decays into bosons 
(Eqs.~(\ref{eq:wzmodes}, \ref{eq:hxmodes})) with those of the 
decays into fermions (Eq.~(\ref{eq:fmodes})). 
The bosonic decays always produce cascade decays. 
In principle, the final states of the bosonic decays 
can also be generated from fermionic decays. 
For example, the final particles of the decay chain 
\beq \label{chaina}
\sa_2^-\to\sa_1^- + (h^0, H^0, A^0\; {\rm or}\; Z^0)\to 
(\tau^- \nt_1) + (b \bar{b}) 
\eeq
are the same as those of 
\beq \label{chainb} 
\sa_2^- \to \tau^- + \nt_{2,3,4} \to 
\tau^- + ((h^0, H^0, A^0\; {\rm or}\; Z^0) + \nt_1) \to 
\tau^- + (b \bar{b} \nt_1). 
\eeq
Likewise, 
\beq \label{chainc}
\sn_\tau \to \sa_1^- + (H^+\; {\rm or}\; W^+) \to 
(\tau^- \nt_1) + (q \bar{q'}) 
\eeq
has the same final particles as 
\beq \label{chaind} 
\sn_\tau \to \tau^- + \chp_{1,2} \to 
\tau^- + ((H^+\; {\rm or}\; W^+) + \nt_1) \to 
\tau^- + (q \bar{q'} \nt_1). 
\eeq
Nevertheless, the decay distributions 
of the two processes (\ref{chaina}) and (\ref{chainb}) 
((\ref{chainc}) and (\ref{chaind})) are in general different from 
each other due to the different intermediate states. 
For example, the $\tau^-$ in the chains (\ref{chaina}) and (\ref{chainc}) 
tends to be softer than the $\tau^-$ in (\ref{chainb}) and (\ref{chaind}), 
respectively. A similar argument holds for the quark pairs 
in the decay chains. 
Moreover, the distribution of the missing energy-momentum carried by 
$\nt_1$ could be significantly different in (\ref{chaina}) and (\ref{chainb}) 
((\ref{chainc}) and (\ref{chaind})) since it is emitted from a different 
sparticle. 
Detailed Monte Carlo simulations are necessary to investigate the 
experimental consequences of the bosonic decays. 


In conclusion, 
we have shown that the decays of $\sa_2$ and $\sn_\tau$ into 
Higgs or gauge bosons, such as 
$\sa_2^-$ $\to$ $\sa_1^-$ + $(h^0,H^0,A^0$~or~$Z^0)$ and 
$\sa_2^-$ $\to$ $\sn_\tau$ + $(H^-$~or~$W^-)$, 
can be very important 
in a fairly wide MSSM parameter region with large mass splitting between 
$\sa_1$ and $\sa_2$, large $\tan\beta$, and large $|A_\tau|$ and/or $|\mu|$. 
Compared to the fermionic decay modes, 
these bosonic decay modes could have significantly different 
decay distributions. This could have important implications for  
the searches of $\sa_2$ and $\sn_\tau$ and 
the determination of the MSSM parameters at future colliders.

\section*{Acknowledgements}

The work of A.B., H.E., S.K., W.M., and W.P. was supported by 
the ``Fonds zur F\"orderung der wissenschaftlichen Forschung'' 
of Austria, project no. P10843--PHY and P13139--PHY.
The work of T.K. and Y.Y. was supported in part by the 
Grant--in--aid for Scientific Research from the Ministry of Education, 
Science, and Culture of Japan, Nos.~08640388 and 10740106, respectively. 
Y.Y. was also supported in part by Fuju--kai Foundation.


\newpage



\begin{flushleft}
{\Large \bf Figure Captions} \\
\end{flushleft}

\noi
{\bf Figure 1}: 
Branching ratios of $\sa_2$ and $\sn_\tau$ decays 
in the $A_\tau$--$\mu$ plane for 
$m_{\sa_1} = 250$ GeV, $m_{\sa_2} = 500$ GeV
and $\tan\beta = 30$ in the cases of 
$m_{\sa_L} < m_{\sa_R}$ (a -- c) and 
$m_{\sa_L} \geq m_{\sa_R}$ (d -- i) ; 
(a, d)
$\sum {\rm BR} 
\big[\: \sa_2^-\to \sa_1^- + (h^0,H^0,A^0),~ 
\sn_\tau + H^- \,\big]$,
(b, e) 
$\sum {\rm BR} 
\big[\: \sa_2^-\to \sa_1^- + Z^0,~ 
\sn_\tau + W^- \,\big]$,
(c, f) 
$\sum {\rm BR} 
\big[\: \sa_2^-\to \sa_1^- + (h^0,H^0,A^0,Z^0),~ 
\sn_\tau + (H^-,W^-) \,\big]$,
(g)
${\rm BR} 
\big[\: \sn_\tau\to \sa_1^- + H^+ \,\big]$, 
(h)
${\rm BR} 
\big[\: \sn_\tau\to \sa_1^- + W^+ \,\big]$, 
and
(i)
$\sum {\rm BR} 
\big[\: \sn_\tau\to \sa_1^- + (H^+,W^+) \,\big]$.
The gray areas are excluded by the conditions 
(i) to (iii) given in the text.

\noi 
{\bf Figure 2}: 
$\tan\beta$ dependence of $\sa_2$ (a, b) and $\sn_\tau$ (c) decay 
branching ratios for 
$m_{\sa_1} = 250$ GeV, $m_{\sa_2} = 500$ GeV, 
$A_\tau = 800$ GeV 
and $\mu = 1000$ GeV in the cases of 
$m_{\sa_L} < m_{\sa_R}$ (a) and $m_{\sa_L} \geq m_{\sa_R}$ (b, c). 
"Gauge/Higgs $+$ $X$" refers to the sum of the gauge and Higgs boson modes. 
The gray areas are excluded by the conditions (i) to (iii) given in the 
text.

\noi 
{\bf Figure 3}: 
$m_{\sa_2}$ dependence of $\sa_2$ (a, b) and $\sn_\tau$ (c) 
decay branching ratios for 
$m_{\sa_1} = 250$ GeV, $A_\tau = 800$ GeV, $\mu = 1000$ GeV, and 
$\tan\beta = 30$ in the cases of 
$m_{\sa_L} < m_{\sa_R}$ (a) and $m_{\sa_L} \geq m_{\sa_R}$ (b, c).
"Gauge/Higgs $+$ $X$" refers to the sum of the gauge and Higgs boson modes. 
The gray areas are excluded by the conditions (i) to (iii) given in the 
text. Note that $m_{\sn_\tau} \sim m_{\sa_2}$ in (c).

\newpage
%
%
%
\begin{figure}[!htb] 
\begin{center}
\scalebox{0.5}[0.5]{\includegraphics{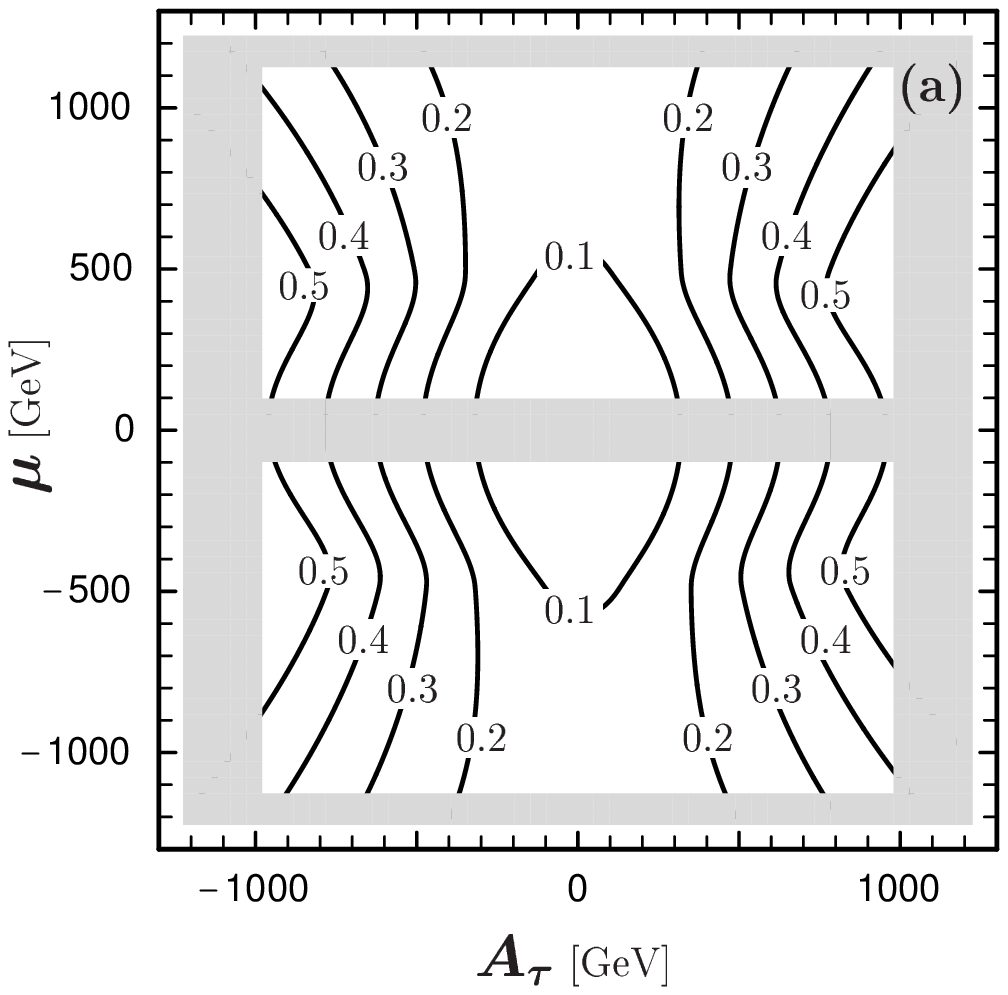}} \hspace{-3.5mm} 
\scalebox{0.5}[0.5]{\includegraphics{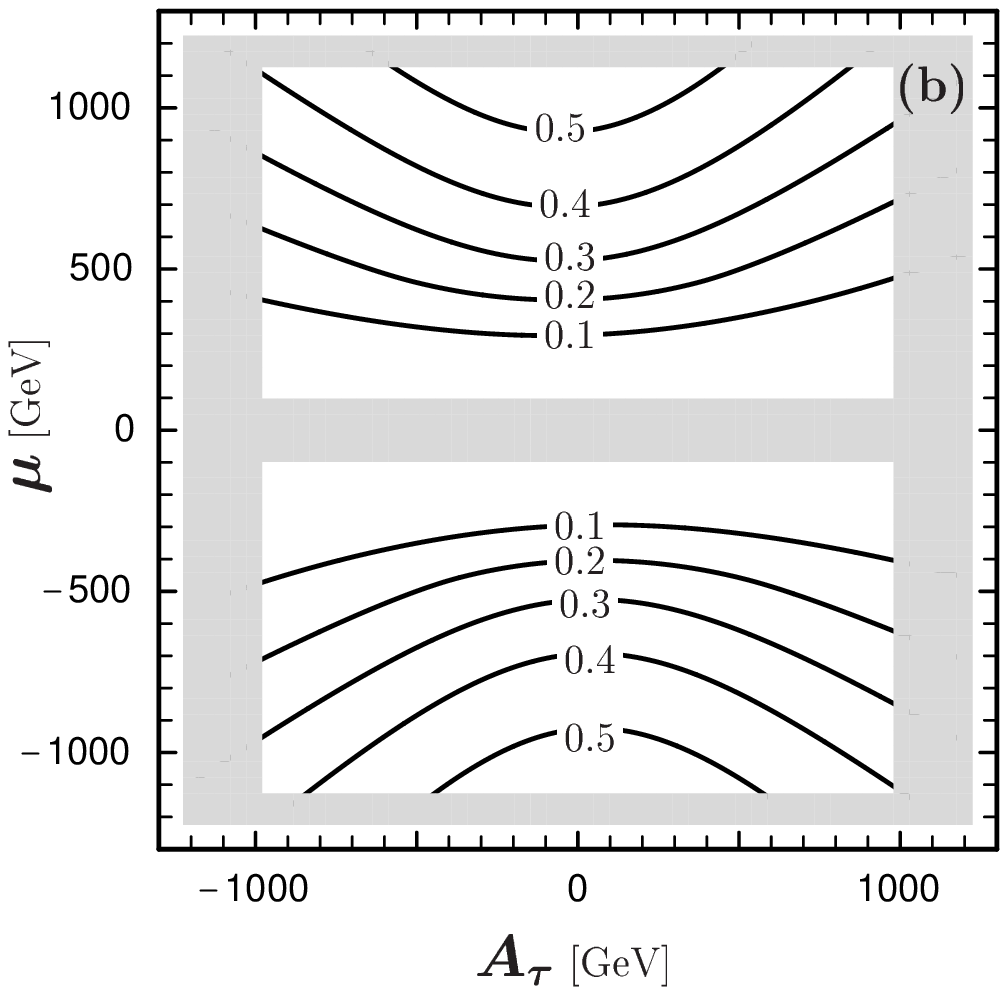}} \hspace{-3.5mm} 
\scalebox{0.5}[0.5]{\includegraphics{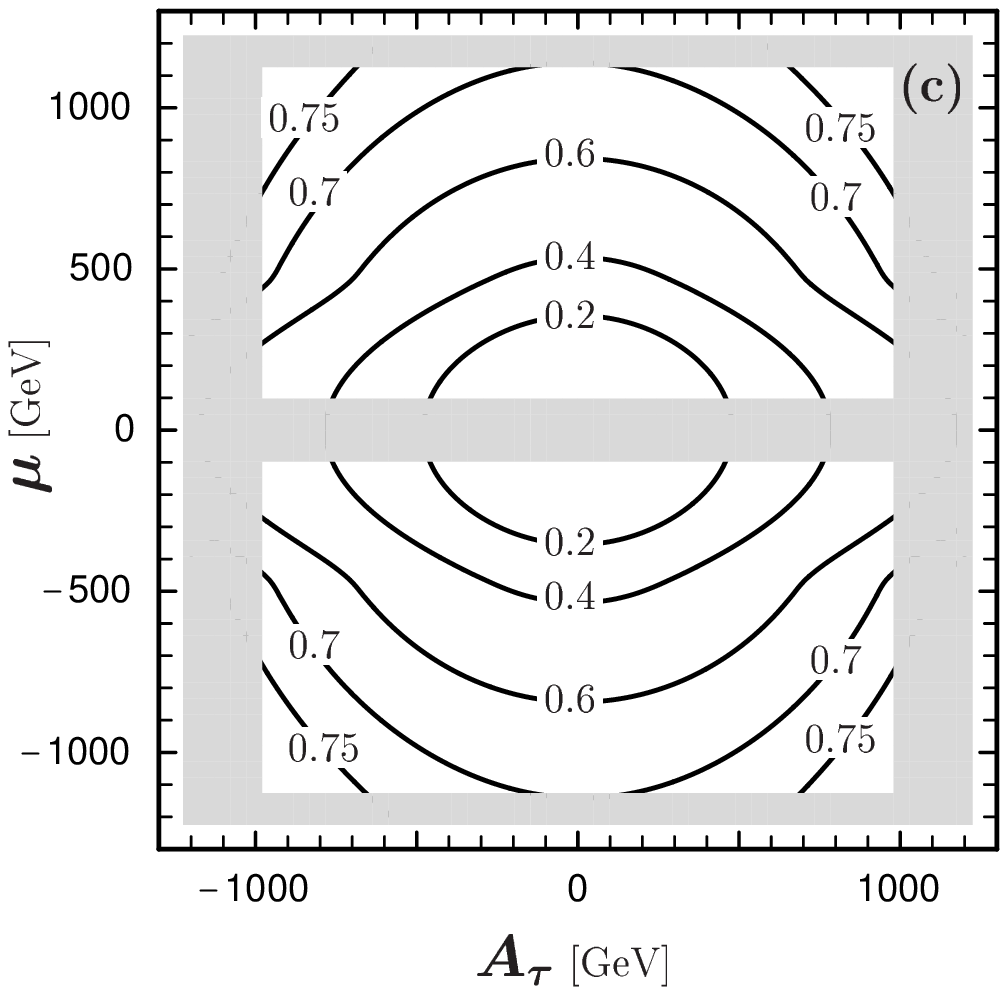}} \\ 
\scalebox{0.5}[0.5]{\includegraphics{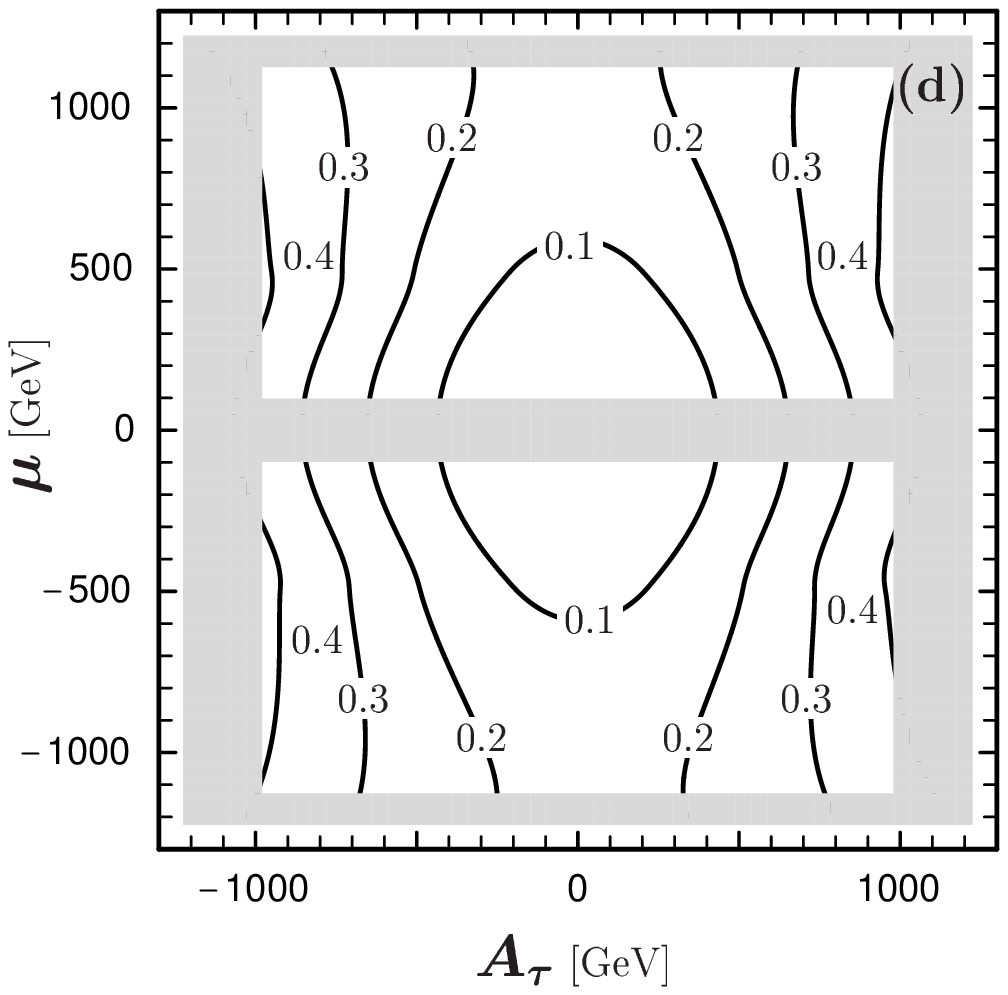}} \hspace{-3.5mm} 
\scalebox{0.5}[0.5]{\includegraphics{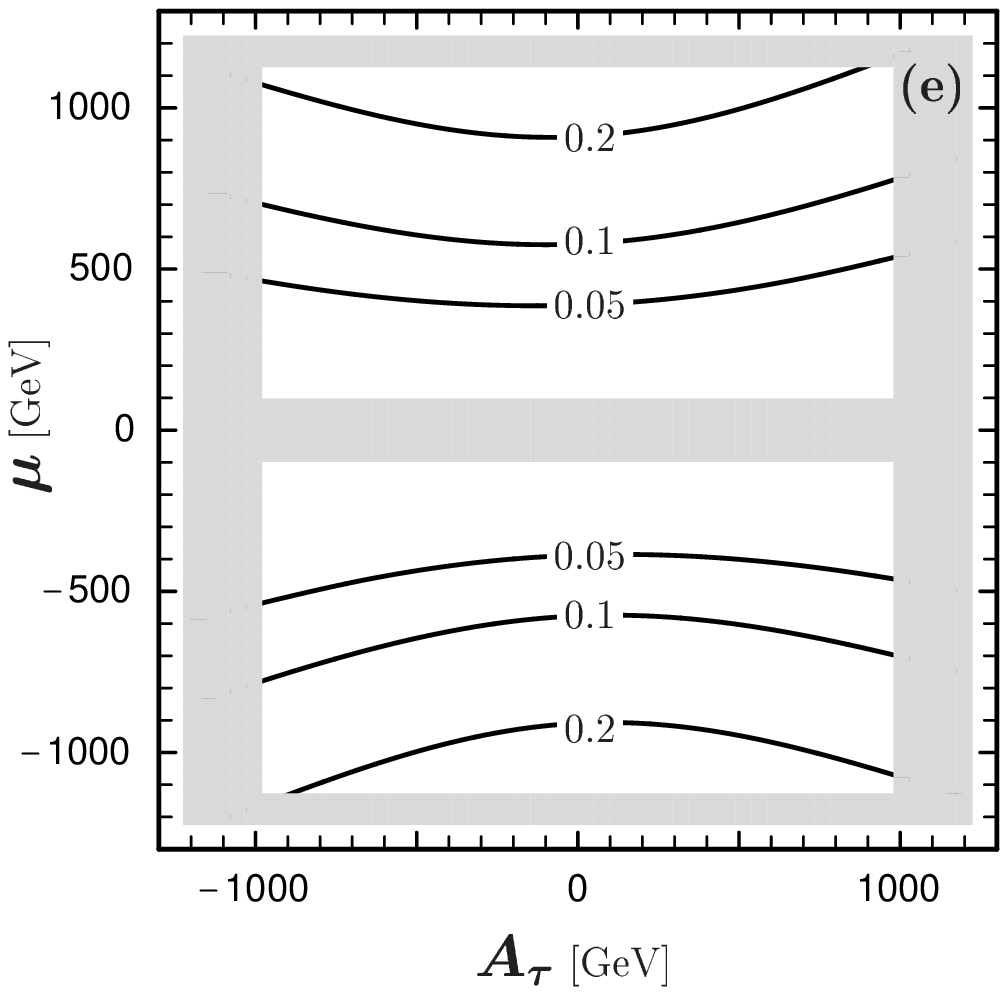}} \hspace{-3.5mm} 
\scalebox{0.5}[0.5]{\includegraphics{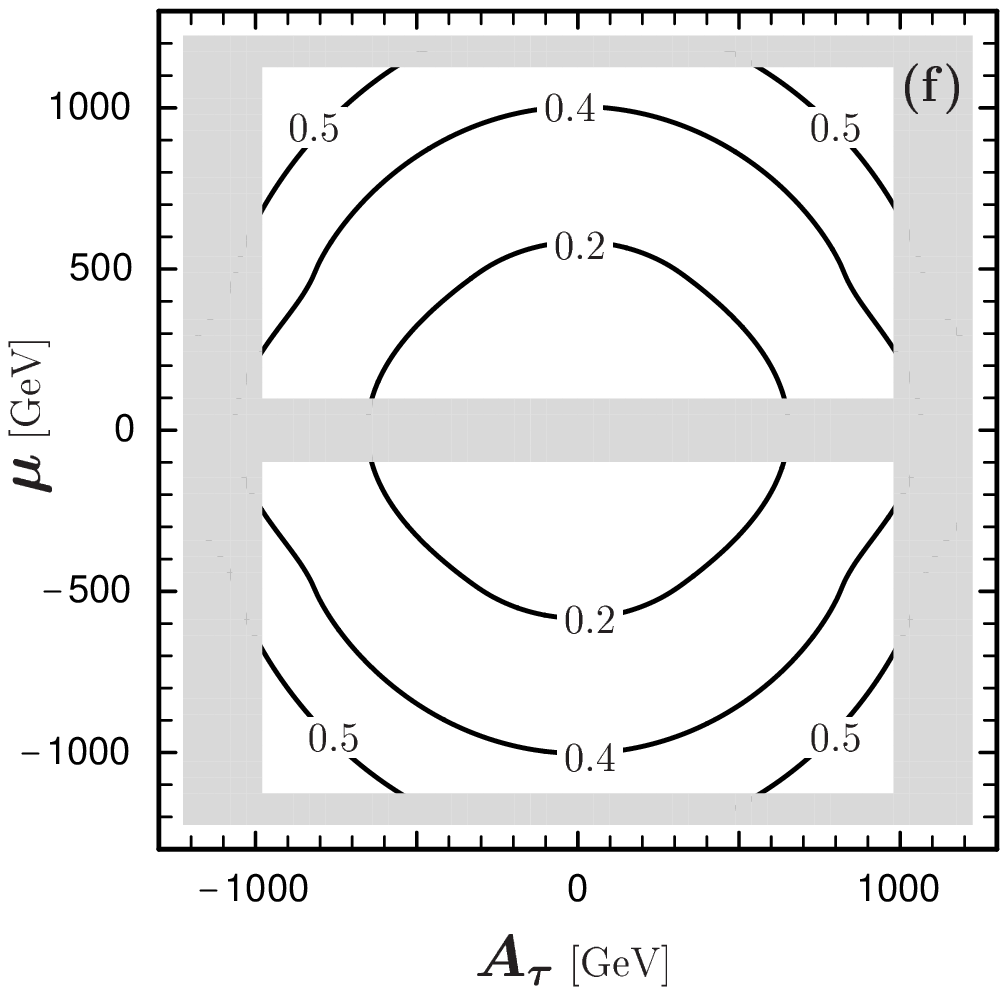}} \\ 
\scalebox{0.5}[0.5]{\includegraphics{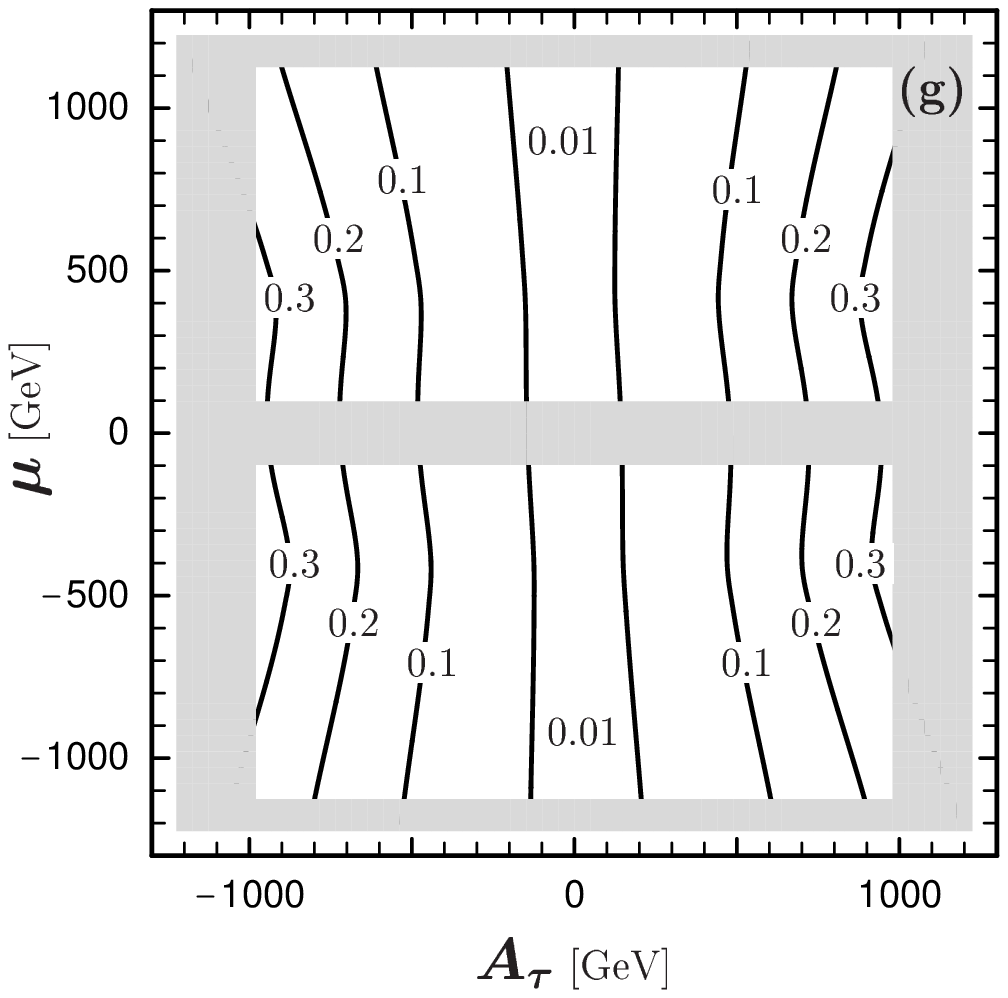}} \hspace{-3.5mm} 
\scalebox{0.5}[0.5]{\includegraphics{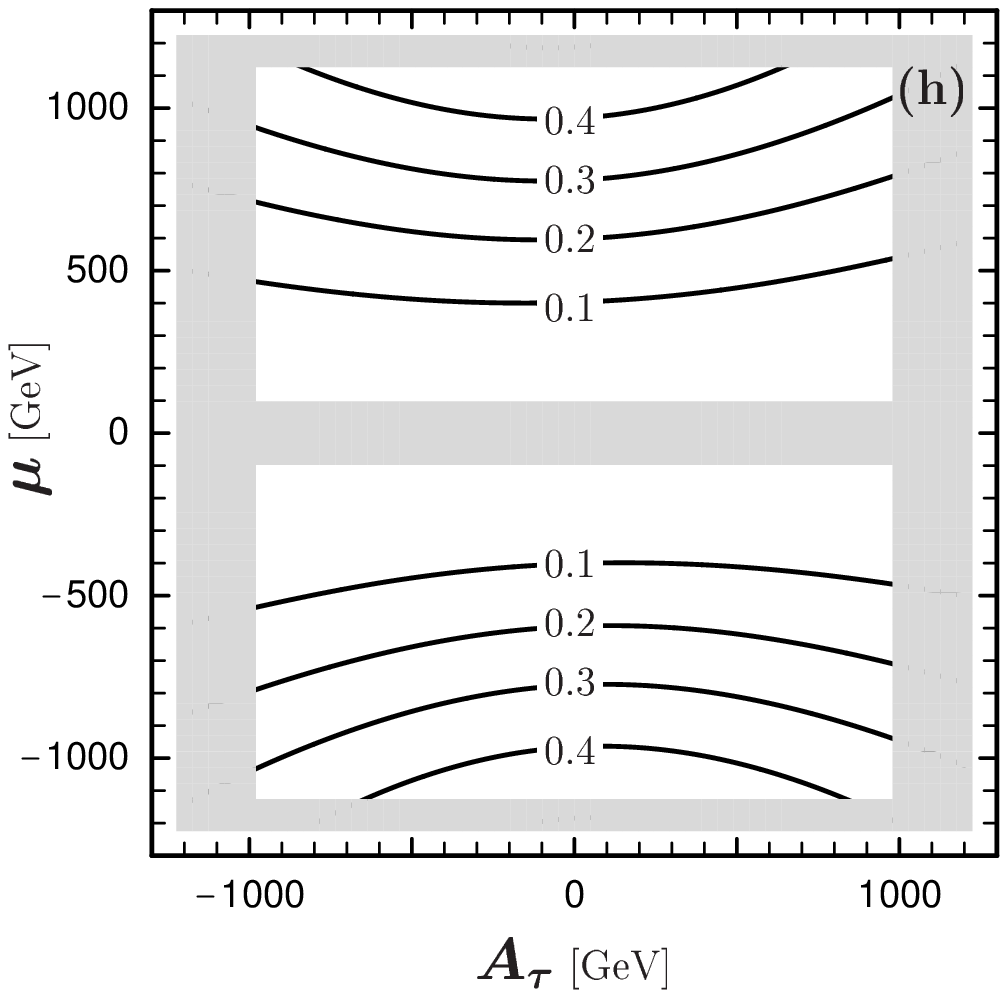}} \hspace{-3.5mm} 
\scalebox{0.5}[0.5]{\includegraphics{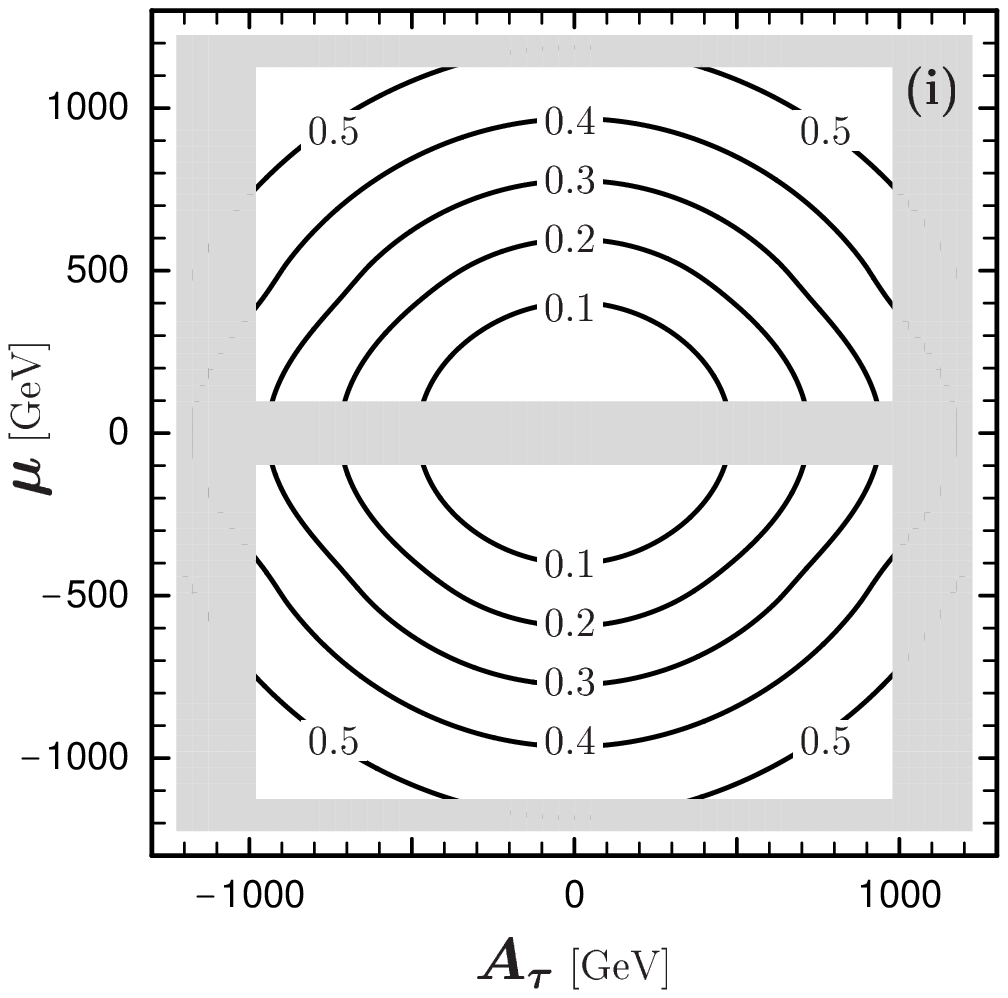}} \\ 
\vspace{5mm}
{\LARGE \bf Fig.1}
\end{center}
\end{figure}
%

\newpage
%
%
\begin{figure}[!htb] 
\begin{center}
\scalebox{0.6}[0.5]{\includegraphics{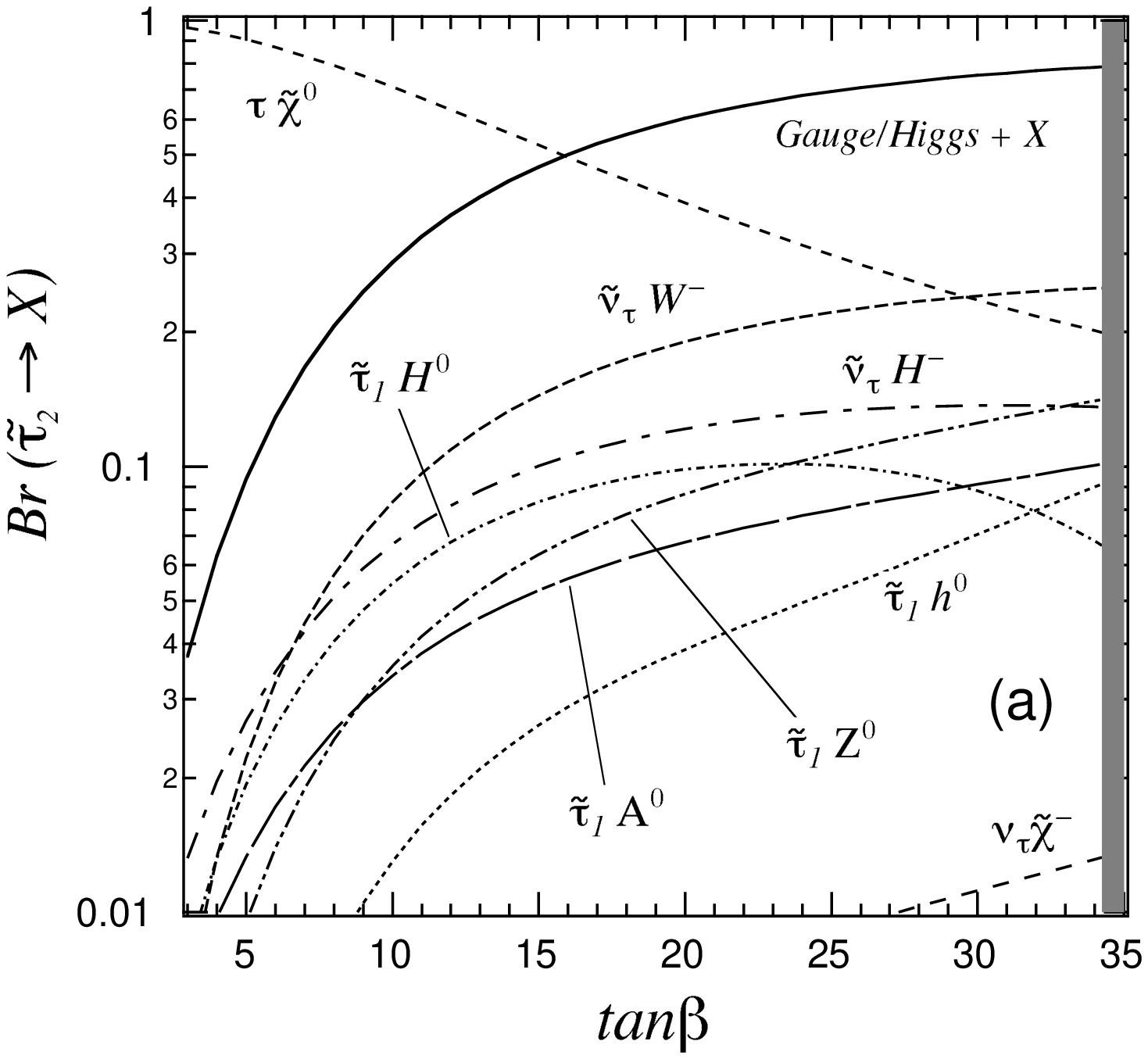}} \\ 
\vspace{1mm}
\scalebox{0.6}[0.5]{\includegraphics{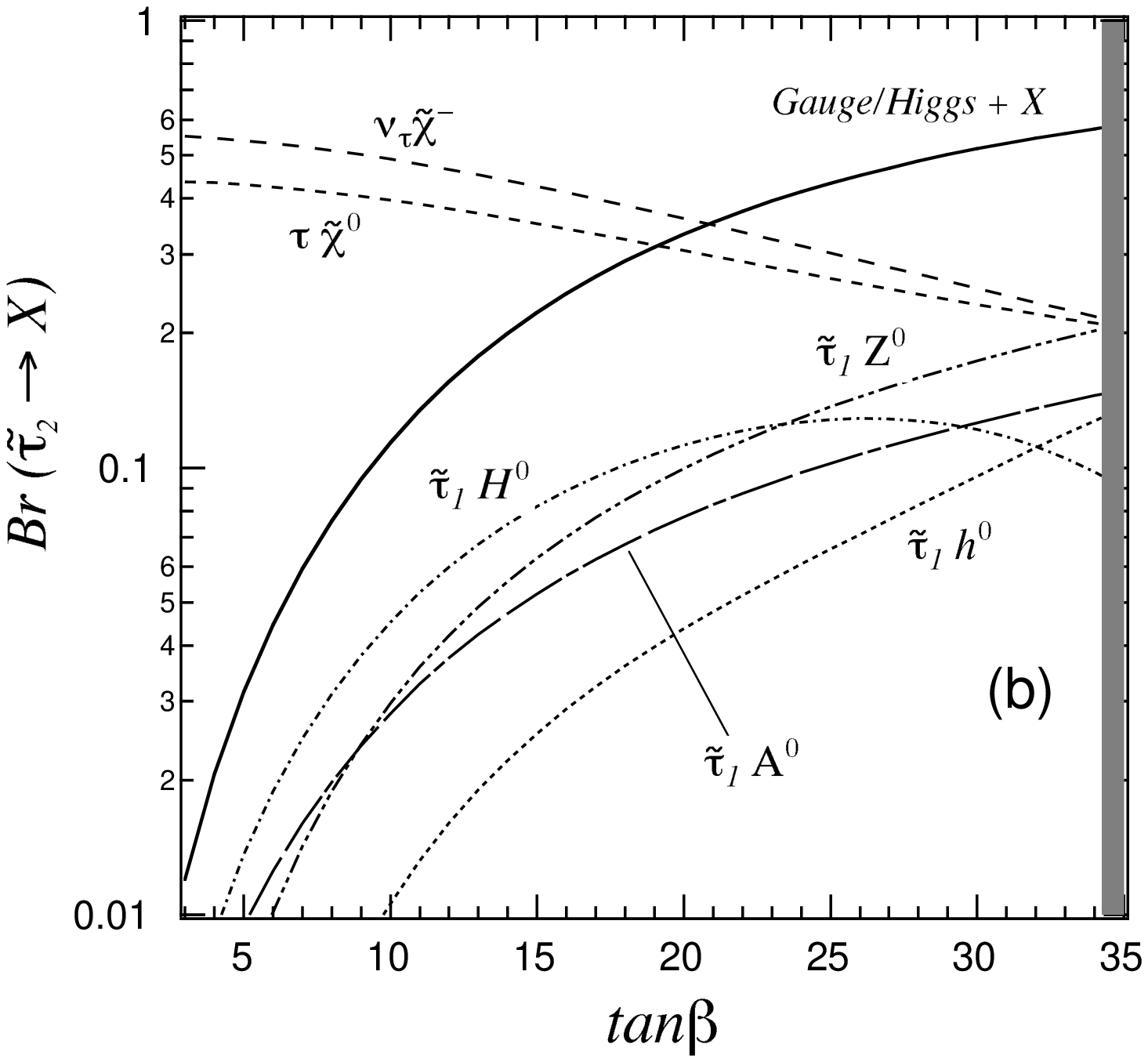}} \\ 
\vspace{1mm}
\scalebox{0.6}[0.5]{\includegraphics{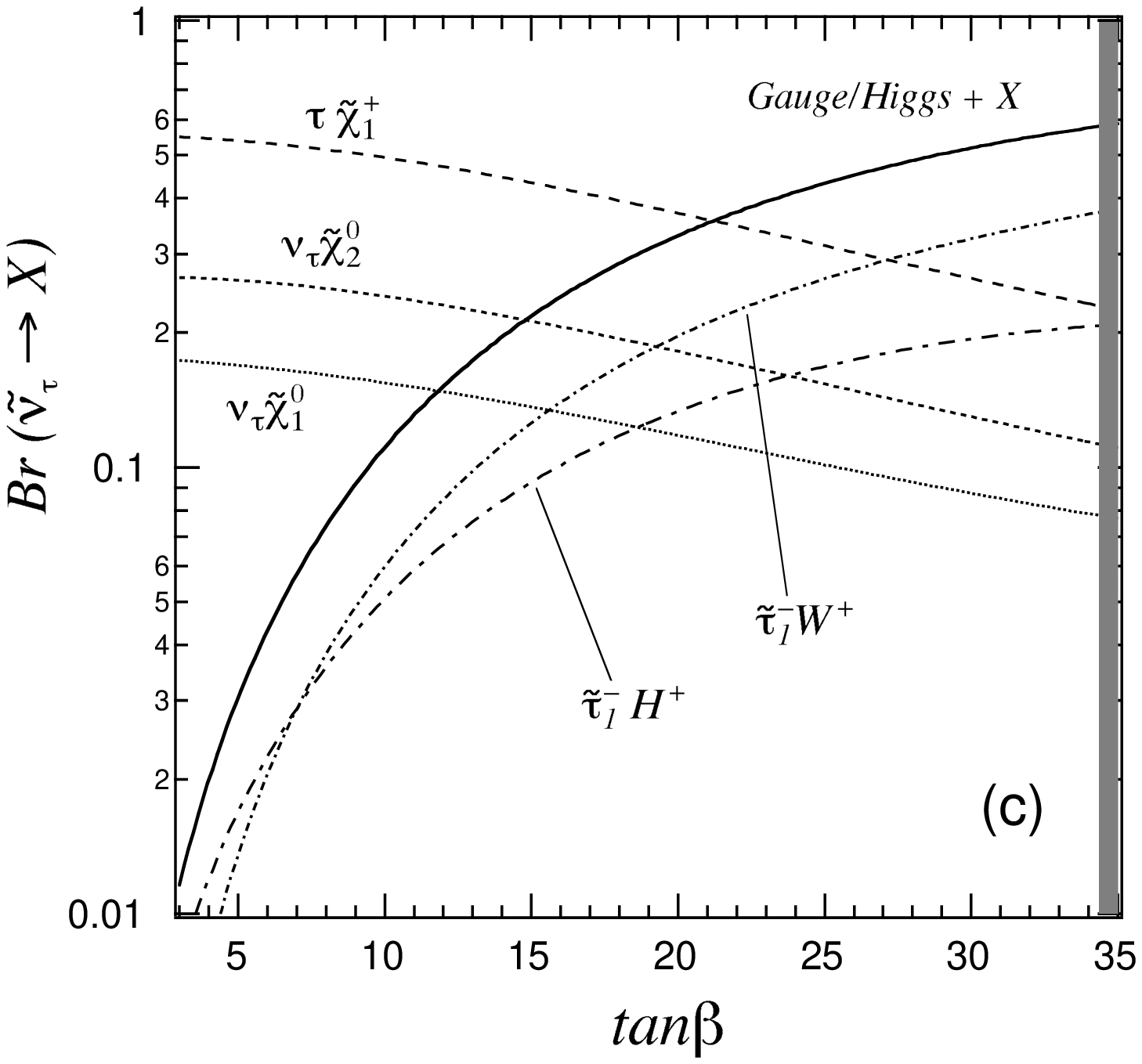}} \\ 
\vspace{5mm}
{\LARGE \bf Fig.2}
\end{center}
\end{figure}
%

\newpage
%
%
\begin{figure}[!htb] 
\begin{center}
\scalebox{0.6}[0.5]{\includegraphics{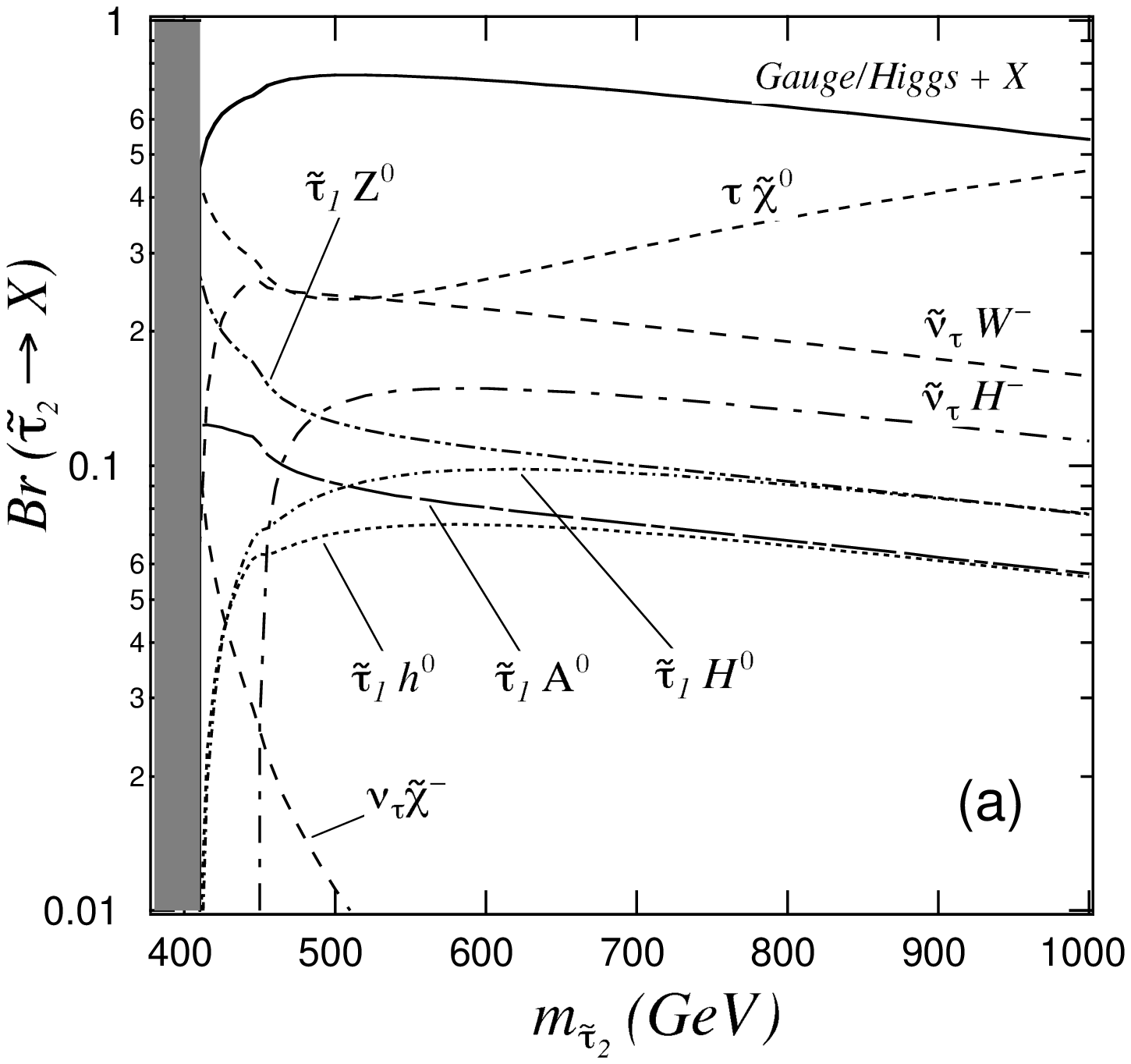}} \\ 
\vspace{1mm}
\scalebox{0.6}[0.5]{\includegraphics{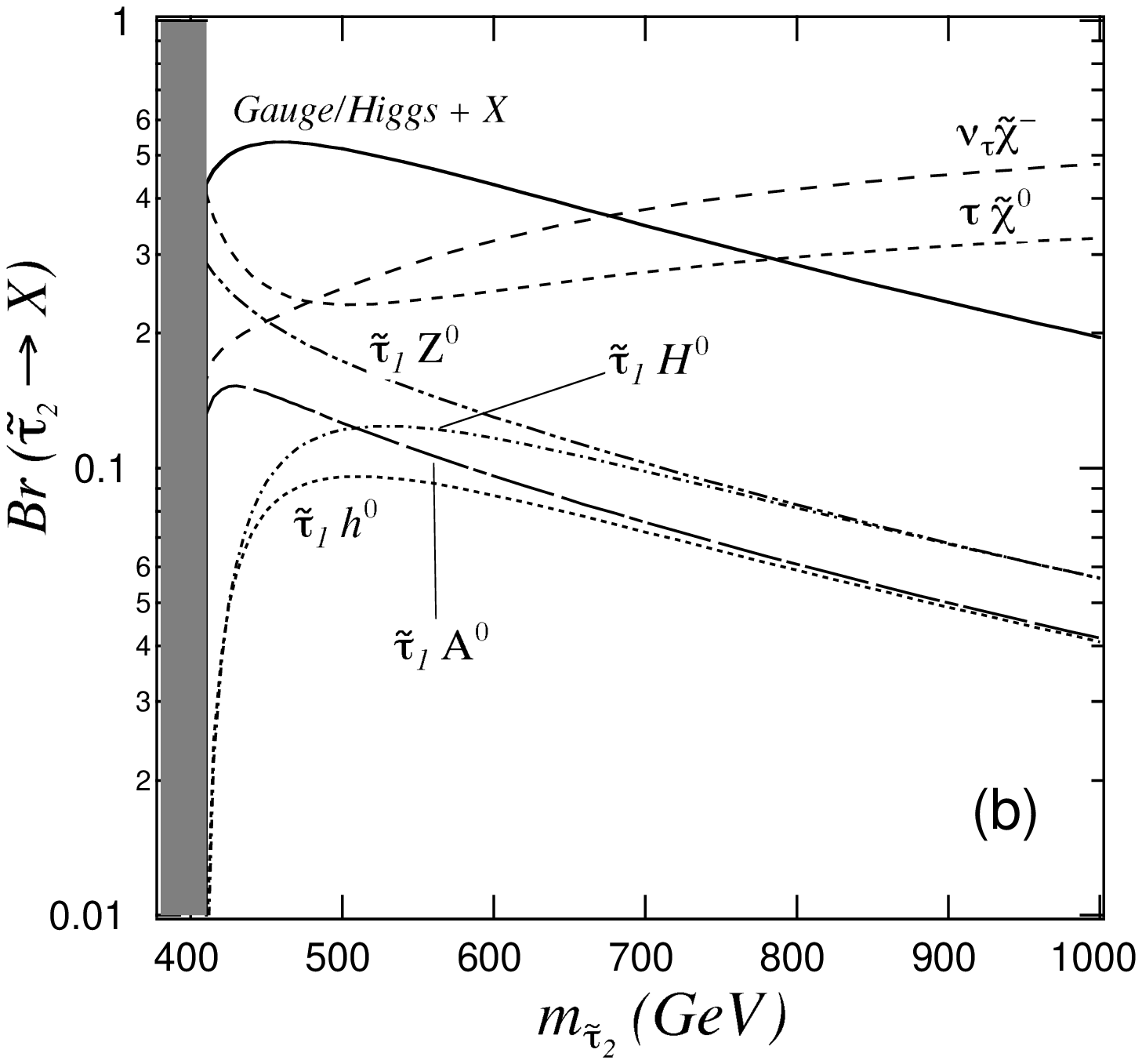}} \\ 
\vspace{1mm}
\scalebox{0.6}[0.5]{\includegraphics{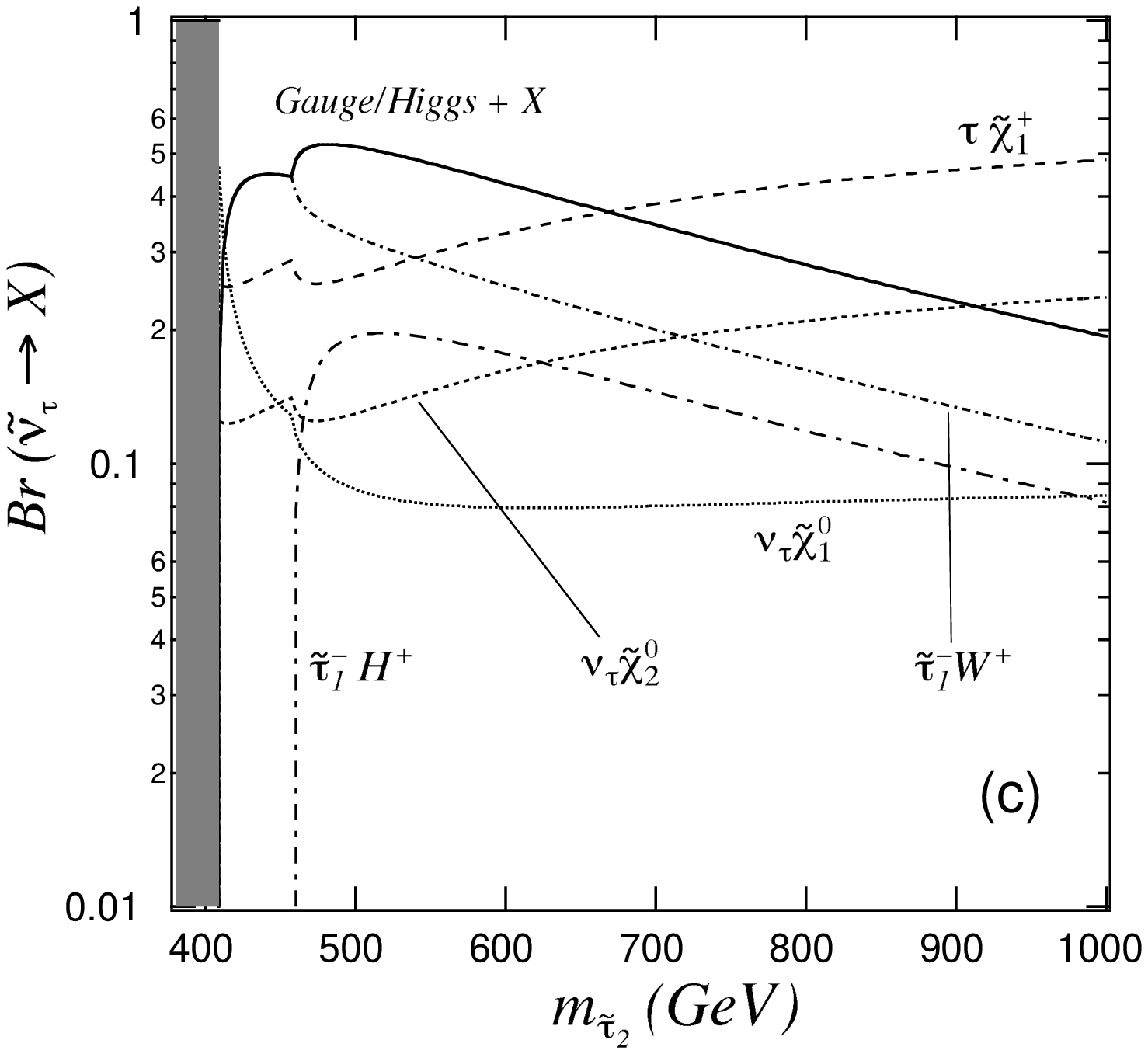}} \\ 
\vspace{5mm}
{\LARGE \bf Fig.3}
\end{center}
\end{figure}
%


\end{document}